# Breast Cancer Classification Using: Pixel Interpolation

**Osama Rezq Shahin\*, Hamdy Mohammed Kelash, Gamal Mahrous Attiya and Osama Slah Farg Allah**

*Department of Computer Sciences, Jouf University, Saudi Arabia*

**\*Corresponding Author:** Osama Rezq Shahin, Department of Computer Sciences, Jouf University, Saudi Arabia.

## Abstract

Image Processing represents the backbone research area within engineering and computer science specialization. It is promptly growing technologies today, and its applications founded in various aspects of biomedical fields especially in cancer disease. Breast cancer is considered the fatal one of all cancer types according to recent statistics all over the world. It is the most commonly cancer in women and the second reason of cancer death between females. About 23% of the total cancer cases in both developing and developed countries. In this work, an interpolation process was used to classify the breast cancer into main types, benign and malignant. This scheme dependent on the morphologic spectrum of mammographic masses. Malignant tumors had irregular shape percent higher than the benign tumors. By this way the boundary of the tumor will be interpolated by additional pixels to make the boundary smoothen as possible, these needed pixels is proportional with irregularity shape of the tumor, so that the increasing in interpolated pixels meaning the tumor goes toward the malignant case. The proposed system is implemented using MATLAB programming and tested over several images taken from the Mammogram Image Analysis Society (MIAS) image database. The MIAS offers a regular classification for mammographic studies. The system works faster so that any radiologist can take a clear decision about the appearance of calcifications by visual inspection.

## Introduction

Breast cancer appears as dense regions of different sizes they can be circular, oval, lobular, or irregular/spiculated (Morphologic spectrum). Mammogram screening was considering the most common and effective method for detecting breast cancer in their early stages [1- 5]. It can indicate potential clinical problems, such as the: architectural distortion, breasts asymmetries, that associated with benign fibrosis, Microcalcification Clusters (MCCs) and mass lesions. The mainly two common features that are typically associated with breast cancer are MCCs and mass lesions.

## Calcifications

Breast calcifications are calcium deposits that form in the tissue of a woman's breast.

It's commonly thought to be a good thing (noncancerous). Certain types of breast calcifications, on the other hand, may be a sign of early breast cancer in some circumstances. Microcalcifications and macrocalcifications are the two types of breast calcifications that will be discussed in the following paragraphs.

**Microcalcification**

Also it is considered as tiny calcium deposits that appear as figure 1 shows white spots on a mammography (breast X-ray). Microcalcifications aren't always a sign of malignancy. However, if they appear in specific patterns, they could be a symptom of early breast cancer [6,7].

**Macrocalcification**

On a mammography, they appear as huge white specks that are typically isolated at random within the breast. It affects about half of all women over the age of 50, and one in every 10 women under the age of 50. Macrocalcifications considered as a noncancerous condition.

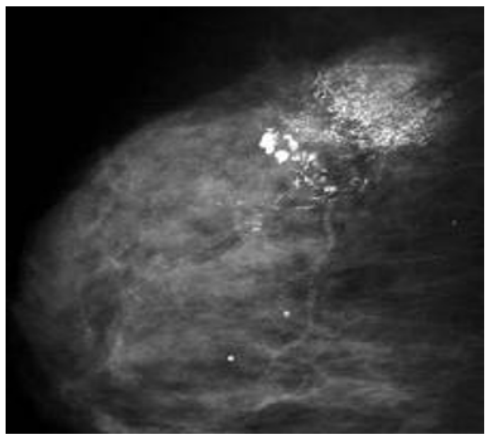

**Figure 1:** Microcalcification clusters (MCCs) in a breast tissue.

Breast cancer is the leading cause of death for women among all cancers [8]. Breast cancer (carcinoma) is a malignant tumor caused by uncontrolled cell division.

**Types of breast tumors**

Although there are many different forms of breast abnormalities, we discriminate between benign and malignant breast abnormalities in this study. The border forms of benign and malignant breast tumors with the surrounding breast tissue can be used to make a general distinction [9,10]. Examining spiculations on the malignant tumor, which can be easily seen utilizing mammography or ultrasound techniques, can help with this differentiation. Spiculation is a distortion caused by breast cancer infiltration into nearby tissue, and it's crucial for identifying the tumor as cancerous as shown in figure 2. Masses can look as thick areas with a variety of features.

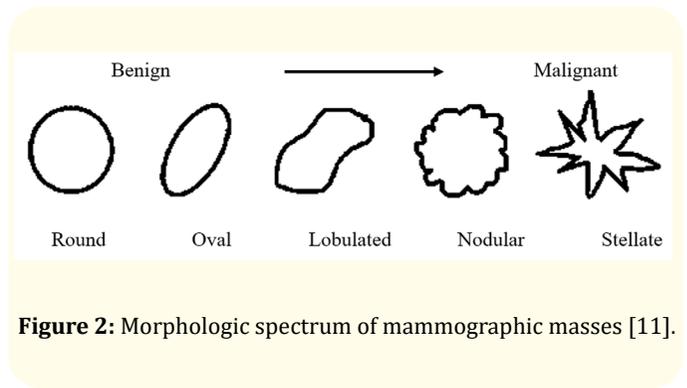

**Figure 2:** Morphologic spectrum of mammographic masses [11].

Although there are many different forms of breast abnormalities, we discriminate between benign and malignant breast abnormalities in this study. The border forms of benign and malignant breast tumors with the surrounding breast tissue can be used to make a general distinction. Examining spiculations on the malignant tumor, which can be easily seen utilizing mammography or ultrasound techniques, can help with this differentiation. Spiculation is a distortion caused by breast cancer infiltration into nearby tissue, and it's crucial for identifying the tumor as cancerous. Masses can look as thick areas with a variety of features [12,13]. Because it's difficult for radiologists to tell the difference between benign and malignant tumors on mammograms, numerous recent research have proven that quantitative methods can help radiologists determine the type of tumor [14]. Figures 3 (a) and (b) depict a benign and malignant tumor, respectively, as examined by [15], and Figures 3 (c) and (d) depict a benign and malignant tumor, respectively, as analyzed by [14].

Other methods separated the tissue tumor boundary into three pieces, each of which was evaluated separately to determine whether the mass lesion was benign or malignant. This is done by looking at the shape of the interface or using mathematical modeling to count the number of jags between the breast tissue and the breast tumor interface to distinguish between benign and malignant tumors [16].

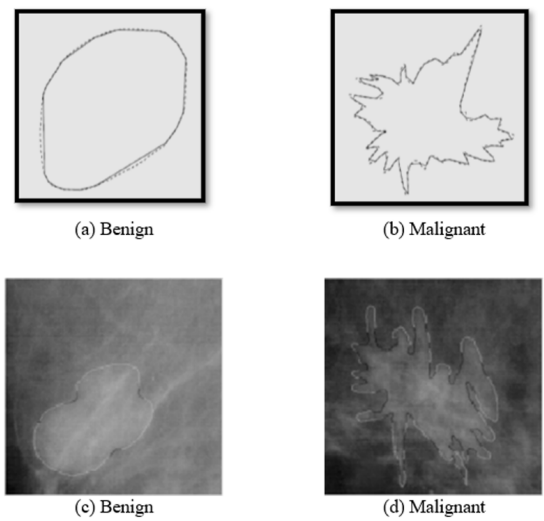

**Figure 3:** Benign and malignant tumors.
Tumors in (a) and (b) are analyzed by [15].
Tumors in (c) and (d) are analyzed by [14].

### BI-RADS descriptors

Radiologists evaluate and offer BI-RADS descriptors, which are crucial determinants in predicting cancers. As shown in figure 2, BI-RADS has developed mass descriptors for shape and margin to detect mass lesions.

- **Shape:** The shape of the mass is defined by a five-point measurement that begins with round, oval, lobular, irregular, and eventually architecturally distorted.
- **Margin:** It denotes the borders of the masses. For example, the mass may appear to be round, but closer study may reveal a succession of erosions along the bulk's usual border.

The geometric features of the discovered tumor can be used to calculate the majority of the above features. In a screening mammography, it is displayed as a cluster of pixels in a specific area. As a result, there is a requirement to illustrate the attributes of pixels [17] for the purpose of identification. The geometric qualities are the fundamental local descriptors that must be recognized by each object. Geometric features such as Area, Perimeter, Circularity, Roundness and Compactness are required in medical analysis to identify the objects located in the medical image, but they are also the key to distinguishing ROI from other normal tissues [18-20].

### Geometrical features for ROI

Area: Is the most basic feature, which determines the tumor's size. As a result, it's the number of pixels in the ROI. Furthermore, the area was unaffected by mass scaling.

The term "area" is defined as:

$$\text{Area} = \sum_i \sum_j (A_{i,j}, X_{ROI(Area)} = i, Y_{ROI(Area)} = j) \quad \text{-------- (1)}$$

Where: i, j are the pixels within the shape. And $X_{ROI}$ is vector contains $ROI_{X\ positions}$, $Y_{ROI}$ is vector contain $ROI_{Y\ positions}$.

### Perimeter

The extent of the extracted ROI boundary defines another simple attribute. Furthermore, it is defined as follows:

$$\text{Perimeter} = \sum_i \sum_j (P_{i,j}, X_{edge(Perimeter)} i, Y_{edge(Perimeter)} = j) \quad \text{------ (2)}$$

Where: $X_{edge}$, $Y_{edge}$ are vector represent the coordinate of the $i^{th}$, $j^{th}$ pixel forming the curve, respectively. Perimeter also is insensitive to scaling and orientation. Where: A is Area and P is perimeter

### Circularity

Is a metric for how round a shape is. The most compact and rounded shape is a circle. The more compact the tumor, the closer it is to being benign in this study. The following equation can be used to determine circularity.

$$\text{Circularity} = \frac{4\pi\ \text{Area}}{\text{Perimeter}^2} \quad \text{--(3)}$$

### Roundness

The roundness of a particle refers to whether or not its edges and corners have been rounded. For a, this is "1." Furthermore, it is provided by.

$$\text{Roundness} = \frac{4\pi\ \text{Area}}{\text{Perimeter}^2} \quad \text{-------(4)}$$

## Compactness

Circles are the most compact shapes. is a property that describes how compact, rounded, or consistent a shape is. The following formula can be used to calculate compactness:

$$\text{Compactness} = \text{Circularity} \times \text{Elongation} \quad\quad\quad (5)$$

Alternatively, the compactness can be given by the following equation:

$$\text{Compactness} = \sqrt{\text{Roundness}} \quad\quad\quad (6)$$

## Methodology

The crucial phase in any CAD system for breast cancer diagnosis is mass classification.

The categorization step of the proposed system is detailed in this section. Following the detection of the tumor, three distinct algorithms were created to extract features from the input mammography picture, with an artificial neural network and a support vector machine being utilized to identify the tumor.

The form of the tumor is an important factor that radiologists consider when examining and extracting questionable spots from mammography. These masses have a morphologic spectrum that resembles a circle [21-23]. The morphologic spectrum of mammographic masses is used in this procedure. Malignant tumors exhibited a larger percentage of irregular shapes than benign tumors. In this method, we interpolate the tumor's boundary by additional pixels to make the boundary as smooth as possible. These needed pixels are proportional to the tumor's irregularity shape, thus as the number of interpolated pixels proportional with irregularity shape of the tumor, so that the increasing in interpolated pixels meaning the tumor goes toward the malignant one.

A diagram for the proposed algorithm is shown in figure 4.

In the following points, the proposed algorithm will be explained in detail.

## Detect region of interests (ROIs)

These regions can be easily detected in an image if the area has sufficient contrast from the background. In this phase, the detection algorithm that discussed in the detection phase was applied. Once the ROI is automatically extracted, the edges detection for the ROI was achieved by Sobel filter. The goal for this step is to isolate this area from the image.

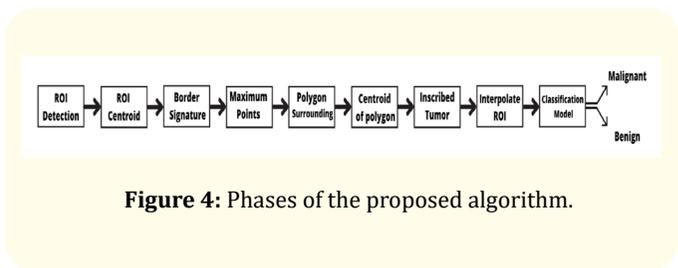

**Figure 4:** Phases of the proposed algorithm.

## Determine centroid for any (ROI)

The centroid of the detected tumor can be calculated from these given equations:

$$\bar{r} = \frac{1}{A} \sum_{(r,c) \in R} r \quad\quad\quad (7)$$

$$\bar{c} = \frac{1}{A} \sum_{(r,c) \in R} c \quad\quad\quad (8)$$

The centroid ($\bar{r}, \bar{c}$) is the average of the pixels location in a region R, Where:

r: Locations of the rows in a region R.

c: Locations of the columns in a region R.

A: Area of the region R.

## Border signature

Calculate the distances from the centroid to all points on the boundary of the ROI as a function of a polar angle θ. A signature is the representation of a 2-D boundary as a 1-D function [3,9]. The signature of a closed boundary is a periodic function, repeating itself on an angular scale of 2π. Such distance called Radial Distance (RD). In figure 5 (a) the object to be detected in the form of circular shape. So, the radial distance remains constant about the object boundary. If the shape is a square or other irregular shape as in figure 5 (b) the radial distance varies with angle θ. A real tumor case is shown in figure 5 (c) with its radial distance. As aforementioned, the importance of RDM to describe the tumor shape is justified.

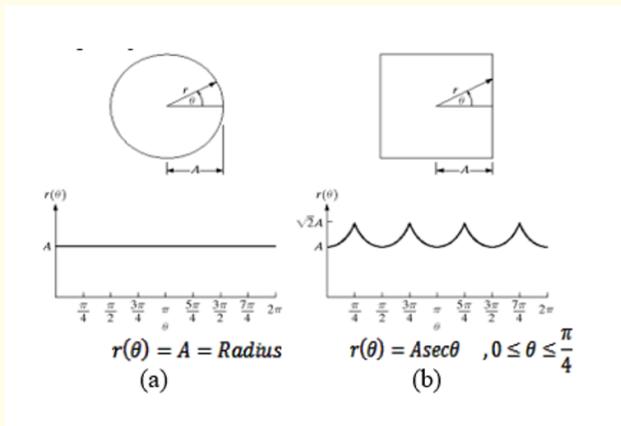

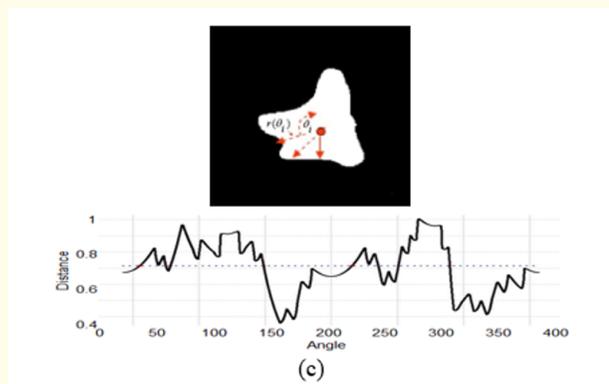

**Figure 5:** Radial Distance Measure (RDM).

**Find the maximum points of the border function.**

Once the border of the region of interest was detected and the function that described the boundary was calculated, the next process is evaluating the second derivative of the signature function, hence the extreme points in that function can be determined as shown in figure 6.

Once the maximum points were determined, the Cartesian coordinate of each point can be calculated from the polar coordinate according to the equations [24]:

$$x_i = r_i \cos\theta_i \quad (9)$$
$$y_i = r_i \sin\theta_i \quad (10)$$

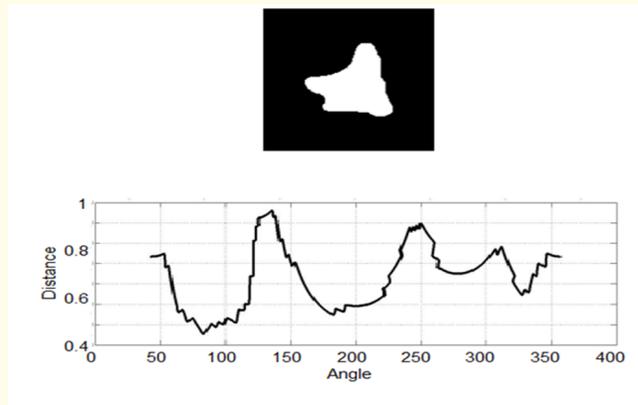

**Figure 6:** Tumor border signature.

Where:

$r_i$: Is the polar distance from the centroid of ROI to the maximum point.

$\theta_i$: Is the angle between the transverse axes to the maximum point.

**Polygon surrounding**

The proposed polygon surround the maximum points calculated from the last step by drawing a straight line between each two successive points. Figure 7 shows this idea.

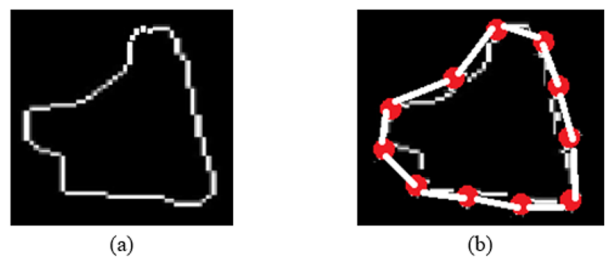

**Figure 7:** Polygon surrounding process, (a) Tumor edge, (b) Polygon between maximum points.

The only constraint that must be sited on the polygon for this method is the lack of the intersections between polygon sides. However, this case cannot occur in this work because only we deal with the boundary of the tumor, which is not self-intersecting.

**Determine the centroid of polygon**

The centroid is the "center of mass". The calculation of the centroid depends on the area calculation of the same polygon. Any polygon prepared from line sectors among N vertices $(x_i, y_i)$, i=0 to N-1. The last vertex $(x_N, y_N)$ is supposed to be similar as the first, i.e.: the polygon is closed [25].

$$A = \frac{1}{2} \sum_{i=0}^{N-1} (x_i y_{i+1} - x_{i+1} y_i) \quad (11)$$

The sign of the last expression of the area can be used to calculate the order of the polygon vertices. If the sign is negative then the vertices of polygon are organized clockwise about the normal, otherwise anticlockwise. After calculation of area, the centroid is calculated by:

$$C_x = \frac{1}{6A} \sum_{i=0}^{N-1} (x_i + x_{i+1})(x_i y_{i+1} - x_{i+1} y_i) \quad (12)$$

$$C_y = \frac{1}{6A} \sum_{i=0}^{N-1} (y_i + y_{i+1})(x_i y_{i+1} - x_{i+1} y_i) \quad (13)$$

**Inscribed tumor polygon in a circle**

Inscribed polygon in a circle is a polygon vertices of which are placed on a circumference. The centre of the circle is the centroid for the proposed polygon and have radius is equal to the average distances between the centroid and all the vertices of the polygon (maximum points), that depicted in figure 8.

**Classification model**

The suggested system's classification model is considered the most important phase. Such a system must be able to efficiently differentiate between benign and malignant. Several automatic learning methods may be used to create such a model.

The differences between the Number of Interpolated pixels in the case of benign and malignant was big enough to use a support vector machine (SVM) to classify between the two classes (Benign, and malignant).

**Experimental Results**

In the first phase of the classification scheme, morphological operations mainly erosion and opening were preformed to remove the small objects in the image. Then the region of interest was detected and the edged image was calculated In order to be able to surround all maximum edged by polygon then surround this polygon by a circular curve. The areas of the objects in the image must be calculated therefore labelling was done to the objects in the image and then the areas were measured. Interpolated the gap between the edged image and the circular curve and determine the number of pixel required for this process which reflected the roundness of the tumor. Figure a shows the output of the proposed algorithm.

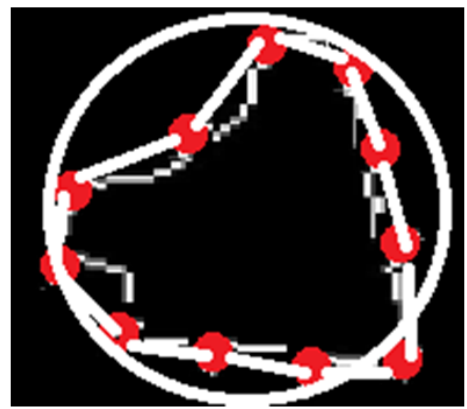

**Figure 8:** Inscribed polygon in a circle.

| Image id | Tumor image | After edge detection | Area of the tumor (pixel) | Image after the interpolation process | Number of Interpolated pixels | Type of the tumor |
|---|---|---|---|---|---|---|
| 1 | | | 836 | | 39 | Malignant |
| 2 | | | 792 | | 44 | Malignant |
| 3 | | | 372 | | 141 | Malignant |
| 4 | | | 603 | | 96 | Malignant |
| 5 | | | 289 | | 5 | Benign |
| 6 | | | 105 | | 2 | Benign |
| 7 | | | 331 | | 12 | Benign |
| 8 | | | 189 | | 10 | Benign |

**Figure a:** Output of the proposed algorithm.

The proposed algorithm is tested on 118 images with various types of masses, including 52 malignant masses and 66 benign masses. The proposed algorithm's confusion matrix is shown in table 1.

Table 2 shows the terminology measurement, of the categorization for the proposed system. The suggested system has a measured accuracy of 94.92%.

| Confusion Matrix | Test Result (Positive or Negative) | |
|---|---|---|
| Disease (Malignant) (True) | TP | FN |
| | 48 | 4 |
| No Disease (False) | FP | TN |
| | 8 | 58 |

**Table 1:** Confusion Matrix.

| Scheme | TN | TP | FP | FN | Sensitivity (%) | Specificity (%) | Accuracy (%) | Precision(%) |
|---|---|---|---|---|---|---|---|---|
| SVM | 61 | 51 | 5 | 1 | 98.07 | 92.42 | 94.92 | 91.07 |

**Table 2**: Terminology measurement of the K-means - Fourier Series.

## Conclusion

Accurate Early discovery means a lower death rate and a higher rate of recovery. As a result, early diagnosis is critical in the treatment of breast cancer. The earlier tumors are detected, the better treatment options are available. The suggested algorithm has a strong relationship with the risk of breast cancer. The SVM model achieved an average classification accuracy of 94.92% in this study. The radiologists will benefit from the system created in this study in a few ways. First, by serving as a second reader after the radiologists, this technology will aid clinical radiologists in the mammographic interpretation process. Secondly, the system is more responsive than most current systems. As a result, any radiologist may make an informed choice based on the appearance of calcifications. When compared to previous investigations, the suggested classifiers are found to be more effective and accurate.